\documentclass[aps,prl,twocolumn,superscriptaddress,floatfix,showpacs]{revtex4}

\bibstyle{apsrev}
\usepackage{amssymb}
\usepackage{graphicx}

\newcommand{\mb}[1]{\mathbf{#1}} 
\newcommand{\etal}{\textit{et al.}}

\begin{document}

\title{Two-band BCS model describes well the thermodynamics of MgB$_2$}

\author{Todor M. Mishonov}
\affiliation{Faculty of Physics, Sofia University "St. Kliment Ohridski", 5
  James Bourchier Boulevard, BG-1164 Sofia, Bulgaria}
\affiliation{Laboratorium voor Vaste-Stoffysica en Magnetisme, Katholieke
  Universiteit Leuven, Celestijnenlaan 200~D, B-3001, Belgium}
\author{Valery L. Pokrovsky}
\affiliation{Department of Physics, Texas A\&M University, College
Station, TX 77843-4242}
\affiliation{Landau Institute for Theoretical Physics,
Chernogolovka, Moscow Distr. 142432, Russia}
\author{Hongduo Wei}
\affiliation{Department of Physics, Texas A\&M University, College
Station, TX 77843-4242}

\begin{abstract}
  Based on weak-coupling anisotropic BCS theory, the temperature
  dependence of energy gap and the specific heat are evaluated for
  MgB$_2$ superconductor, and the results are compared with
  experimental data.  We show that the weak-coupling anisotropic BCS
  theory describes thermodynamic experimental data with high precision
  3--6\%.
\end{abstract}

\maketitle


A keen interest excited by discovery and experimental investigation of
a new high $T_c$ superconductor MgB$_2$ is to a large extent
associated with its dissimilarity to cuprate superconductors. The
superconductivity of MgB$_2$ is definitely three-dimensional effect,
whereas in cuprates it is presumably two-dimensional
(2D). Nevertheless, the superconducting gap in MgB$_2$ displays strong
anisotropy. The most precise tunneling measurements by
Gonnelli~\etal~\cite{gonnelli} give the value $2.6$ for the ratio of
the gaps at two conductivity bands. On the other hand, the measured
gaps are the same for the tunneling in $ab$-plane and in
$c$-direction, indicating that they do not depend on direction within
each piece of the Fermi surface.

An important problem is how strong is the interaction in MgB$_2$.
First-principles calculations~\cite{mazin,golubov,shulga} indicate
that electron-phonon interaction is not weak and that Eliashberg
description is appropriate. However, anisotropy and interaction were
shown to influence thermodynamics oppositely. For example, the
anisotropy decreases the relative discontinuity of the specific heat
at the transition point~\cite{pokrovsky1,pokrovsky2}, whereas the
first correction due to interaction increases
it~\cite{gelikman-kresin}.  Besides, MgB$_2$ is a very hard material
with high value of Debye frequency, which usually correlates with a
weak coupling. Therefore it is not \textit{a priori} clear what is
more substantial in the case of MgB$_2$.

The purpose of our work is to demonstrate that the anisotropy effects
are more substantial at least for thermodynamic measurements. We show
that, as a matter of fact, the weak coupling anisotropic BCS theory
describes all known thermodynamic experimental data including the
temperature dependence of the energy gap and specific heat with a high
precision 3--6\%.

The main features of anisotropic weak coupling BCS model were
elucidated in the early
1960s~\cite{pokrovsky1,pokrovsky2,hohenberg,markowitz,clem}, the
ultimate result being the factorization of the gap~\cite{pokrovsky1}
\begin{equation}
\Delta(T,\mb{k})=Q(T)\chi(\mb{k})\,,
\label{factor}
 \end{equation}
which was experimentally verified by Zavaritskii~\cite{zavaritskii}.
The function of angle $\chi(\mb{k})$ is the eigenfunction of the
interaction operator $V(\mb{k},\mb{k}^{\prime})$ corresponding to the
maximal eigenvalue $\lambda_{+}$. It satisfies linear homogeneous integral
equation:
\begin{equation}
\int V(\mb{k},\mb{k}^{\prime}) \chi(\mb{k}^{\prime}) \frac{d
\sigma^{\prime}} {\nu_F v_F}=\lambda_{+}\, \chi(\mb{k}) \,.
\label{eigen-eq}
\end{equation}
Integration in Eq. (\ref{eigen-eq}) proceeds over the Fermi
surface with $d\sigma=\frac{dS}{8\pi^3}$ and $dS$ being a
differential  area of the Fermi surface; $\nu_F=\int \frac{d
\sigma}{v_F}$ is the electron density of the state per spin at
Fermi level. The function $\chi(\mb{k})$ is normalized as follows:
 \begin{equation}
 \langle \chi^2(\mb{k})\rangle =1
 \label{normalization}
 \end{equation}
The angular average value $\langle X\rangle$ is: $\langle
X\rangle=\int \frac{X d \sigma}{\nu_F}$. the temperature
dependent factor $Q(T)$ can be found from the orthogonality condition:
\begin{equation}
\ln\frac{Q(0)}{Q(T)} =\left< \chi^2(\mb{k})
F\left(\frac{Q(T)\chi(\mb{k})}{T}\right)\right>,
\label{orth-cond}
\end{equation}
where
\begin{equation}
F(x)=\int_{-\infty}^{+\infty}
    \frac{du}{\sqrt{x^2+u^2}(\exp{\sqrt{x^2+u^2}}+1)}\,.
\label{fx}
\end{equation}
The value $Q(0)$ is associated with the transition temperature $T_c$
by the following relationship:
\begin{equation}
\frac{Q(0)}{T_c}=\frac{\pi}{\gamma}\exp(-\langle\chi^2({\bf k})\ln|\chi({\bf k})|\rangle) \,,
\label{Q0-Tc}
\end{equation}
here $\gamma=e^{C}=1.781072\cdots$ and $C$ is Euler's constants.
The specific heat $C(T)$ reads:
\begin{equation}
 C(T)= 2\nu_F T \frac{d}{d T} \left< \Delta_{\mb{k}}\,
 G\left(\frac{\Delta_{\mb{k}}}{T}\right) \right>,
\label{heat-capacity}
\end{equation}
where $G(x)=2x\int_{0}^{\infty}\cosh (2\varphi) F(x\cosh \varphi)d
\varphi$.

We now apply these formulas to MgB$_2$. The Fermi surface of MgB$_2$
has two $\sigma$-type 2D cylindrical hole sheets and two $\pi$-type
three-dimensional tubular networks~\cite{Belashchenko,choi,liu}. We
accept a simple model introduced first by
Moskalenko~\cite{Moskalenko}, in which the interaction does not depend
on the momentum inside each band, but only on the band index. Thus, it
can be written as $2\times 2$ Hermitian matrix $V_{ik}$ ($i,k
=\sigma$, $\pi$). The order parameter (energy gap) in each band in
such a model does not depend either on the momentum within each band
and can be described by a 2D vector with components $\Delta_{\sigma}$,
$\Delta_{\pi}$. The validity of this simple model is supported by the
tunneling measurements of the energy gap~\cite{gonnelli}. which
displays the same values for two gaps in $ab$-plane and in
$c$-direction. The normalized wave function of the Cooper pairs
$\chi_{\mb{k}}$ has the same property:
$\chi_{\sigma}(\mb{k})=\chi_{\sigma}$,
$\chi_{\pi}(\mb{k})=\chi_{\pi}$, where $\chi_{\sigma}$ and
$\chi_{\pi}$ are two constants. We introduce an additional
simplification assuming these constants to be real.  Let us denote the
density of states in the $\sigma$ and $\pi$ bands as $\nu_{F\sigma}$
and $\nu_{F\pi}$, respectively. Then the definition of an average
value $\langle X\rangle$ for any physical value $X$, which does not
change within each band reads:
\begin{equation}
\langle X\rangle=X_{\sigma}c_{\sigma}+X_{\pi}c_{\pi}\,,
\label{twoband-average}
\end{equation}
where $c_{\sigma}$ and $c_{\pi}$ are statistical weights of the bands
$c_{\sigma}=\nu_{F\sigma}/\nu_{F}$ and $c_{\pi}=\nu_{F\pi}/\nu_{F}$,
$\nu_{F}=\nu_{F\sigma}+\nu_{F\pi}$.  The general normalization
condition Eq.~(\ref{normalization}) for this model reads:
\begin{equation}
\chi_{\sigma}^2 c_{\sigma}+\chi_{\pi}^2 c_{\pi}=1\,.
\label{norm-twoband}
\end{equation}
Equation~(\ref{Q0-Tc}) can be written explicitly as follows:
\begin{equation}
\frac{Q(0)}{T_c}=\frac{\pi}{\gamma \chi_{\mathrm{av}}} \,,
\label{Q0-Tc-twoband}
\end{equation}
where $\chi_{\mathrm{av}}=\chi_{\sigma}^{\chi_{\sigma}^2
c_{\sigma}}\chi_{\pi}^{\chi_{\pi}^2 c_{\pi}}$. We assume the values
$c_{\sigma}= 0.44$ and $c_{\pi}= 0.56$ as found from
density-functional theory calculations in
Refs.~\cite{Belashchenko,choi,liu}. The second fitting parameter is
$T_c$. There is no experimental discrepancy on this value, and it is
commonly accepted to be $T_c\approx 39$~K. One additional fitting
parameter for the two-band theory is the ratio $\delta=\chi_{\sigma}/
\chi_{\pi}$. We have extracted it from the tunneling gap
measurements~\cite{{gonnelli}} extrapolating them to zero temperature:
\begin{equation}
\delta=\chi_{\sigma}/ \chi_{\pi}\approx 2.54 \,.
\label{delta}
\end{equation}
Equations~(\ref{Q0-Tc-twoband}) and (\ref{delta}) allow us to
determine $\chi_{\sigma}$ and $\chi_{\pi}$ separately:
$\chi_{\sigma}=\delta/\sqrt{c_{\sigma}\delta^2+c_{\pi}}=1.38$;
$\chi_{\pi}=1/\sqrt{c_{\sigma}\delta^2+c_{\pi}}=0.54$. According to
the weak-coupling theory, the ratio $\delta$ must be the same at any
temperature. This crucial condition is satisfied in the tunneling
experiment~\cite{gonnelli} with all experimental precision.

\begin{figure}[ht]
\includegraphics[width=85mm]{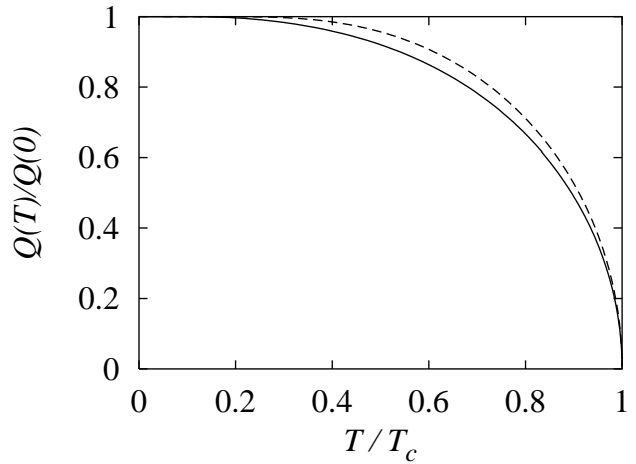}
\caption{\label{f1} The solid curve depicts the ratio $Q(T)/Q(0)$
vs $t=T/T_c$ for the two-band model; the dashed curve is the same
value for the standard (isotropic) BCS theory.}
\end{figure}
\begin{figure}[ht]
\includegraphics[width=85mm]{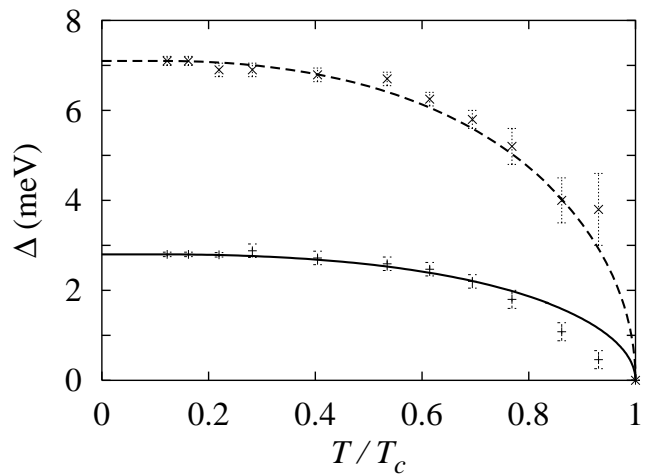}
\caption{\label{f2} The solid curve is the theoretical graph of
$\Delta_{\sigma}$ vs $T/T_c$; the dashed curve is the same for
$\Delta_{\pi}$; ``$+$'' and ``$\times$'' represent experimental data
by Gonnelli~\etal~\protect\cite{gonnelli}}
\end{figure}

For the temperature dependence of the gap in the BCS two-band model,
we find from Eq.~(\ref{orth-cond}):
\begin{equation}
-\ln{q}=\chi^2_{\sigma}F\left(\frac{\pi \chi_{\sigma} q}{\gamma
 \chi_{\mathrm{av}} t}\right)c_{\sigma}
 +\chi^2_{\pi}F\left(\frac{\pi\chi_{\pi} q}{\gamma \chi_{\mathrm{av}}
 t}\right)c_{\pi}\,.
\label{q-t}
\end{equation}
Here $q(t)=Q(t)/Q(0)$ and $t=T/T_c$. $F(x)$ is defined by
Eq. (\ref{fx}). The graph of the function $q(t)$ is shown in
Fig.~\ref{f1} by the solid curve. The dashed curve in Fig.~\ref{f1}
represents $q(t)$ in the isotropic single-gap model (standard BCS
model). The graphs of the energy gaps
$\Delta_{\sigma}=Q(t)\chi_{\sigma}$ and $\Delta_{\pi}=Q(t)\chi_{\pi}$
vs $T/T_c$ are shown in Fig.~\ref{f2} together with the experimental
data~\cite{gonnelli}, which agree with theory within the limits of
experimental uncertainty.

The specific heat in the two-band model is given by the following
equation directly stemming from Eq.~(\ref{heat-capacity}):
\begin{eqnarray}
\frac{C(T)}{C_N(T)}&=&c_{\sigma}r_{c}(y_\sigma)+c_{\pi}r_{c}(y_\pi)
\nonumber \\
&+&\frac{12}{7\zeta(3)}\frac{[c_{\sigma}\chi_{\sigma}^2
r_{a}(y_\sigma)+c_{\pi}\chi_{\pi}^2 r_{a}(y_\pi)]^2}
{c_{\sigma}\chi_{\sigma}^4 r_{b}(y_\sigma)+c_{\pi}\chi_{\pi}^4
r_{b}(y_\pi)},
\label{specialheat}
\end{eqnarray}
where $C_{N}(T)=\gamma T$ is the specific heat for the normal metal;
$y_{\sigma}=\frac{\pi}{2
\gamma}\frac{q}{t}\frac{\chi_{\sigma}}{\chi_{\mathrm{av}}}$,
$y_{\pi}=\frac{\pi}{2
\pi}\frac{q}{t}\frac{\chi_{\pi}}{\chi_{\mathrm{av}}}$.
The functions $r_i$ are defined by integrals
$r_i(x)=\int_{-\infty}^{+\infty}g_i(\sqrt{x^2+y^2})dy$, $i=a$, $b$,
$c$, where $g_i$ read:
\begin{eqnarray}
g_a(x) &=& \frac{1}{2\cosh^2 (x)}, \nonumber \\
g_b(x) &=& \frac{\pi^2}{14 \zeta(3)}
       \left( \frac{\tanh x}{x}-\frac{1}{\cosh x} \right)
            \frac{1}{x^2},  \nonumber \\
g_c(x) &=& \frac{6}{\pi^2}\frac{x^2}{\cosh^2 x}.
\end{eqnarray}
For technical details related to this calculation see
Mishonov~\etal~\cite{mishonov2}; the functions $g_i$ were introduced and
graphically presented in Ref.~\cite{mishonov3}
\begin{figure}[t]
\includegraphics[width=85mm]{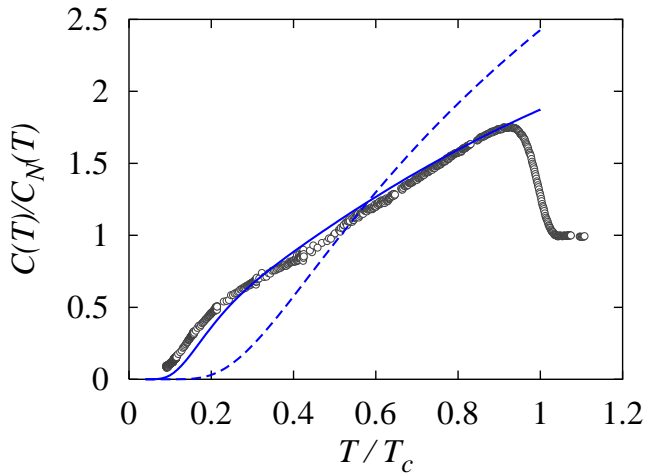}
\caption{\label{f3} The solid curve is the theoretical graph of the
specific heat for the two band MgB$_2$ vs $t=T/T_c$; the circles are
the experimental data due to Bouquet~\etal~\protect\cite{Bouquet}; the
dashed curve is the  theoretical plot of the specific heat given by
the isotropic BCS theory.}
\end{figure}
The jump of the specific heat at $T_c$ reads:
cf.\cite{pokrovsky1,Moskalenko}
\begin{equation}
\frac{\Delta C(T_c)}{C_N(T_c)}=\frac{12}{7\zeta(3)}
 \frac{\left(\chi_{\sigma}^2c_{\sigma}+\chi_{\pi}^2c_{\pi}\right)^2}
{\chi_{\sigma}^4 c_{\sigma}+\chi_{\pi}^4 c_{\pi}}\,.
\label{jump-C}
\end{equation}
For the data specified earlier, we find $\Delta
C(T_c)/C_N(T_c)=0.874$.  It agrees with the high precision
measurements by Bouquet~\etal~\cite{Bouquet} with about $3$\%
precision. In Fig.~\ref{f3} the ratio $C(T)/C_N(T)$ vs. $T/T_c$ is
plotted. The solid curve is the prediction of the two-band weak
coupling theory; the dots are experimental data by
Bouquet~\etal~\cite{Bouquet}, courteously sent to us by the
authors. The theoretical graph $C(T)/C_{N}$ vs $T/T_c$ agrees well
with the experimental data everywhere except of a range of low
temperature $T/T_c \leq 0.2$. The discrepancy most probably is caused
by a relatively small variation of the gap within one band. The
specific heat at low temperature is proportional to
$e^{-\Delta_{\mathrm{min}}/T}$, whereas the tunneling measurements
give the value of the gap along the direction of the tunneling. Given
the value of discrepancy, we can estimate the variation of the gap
$\bar{\Delta}-\Delta_{\mathrm{min}} \sim 0.1\mbox{--}0.15\,T_c \ln
2\approx 3.3\mbox{--}4.2$~K. It is about 8--12\% of the value of the
smaller gap.

Another group of available experimental thermodynamic data relates to
magnetic properties: the energy gaps in external magnetic
field~\cite{gonnelli2} and the dependence of the second critical field
on temperature~\cite{lyard}. The dependence of $H_{c2}$ on temperature
was considered theoretically in the framework of anisotropic BCS model
by two groups of authors~\cite{kogan2,dahm}. based on classical
approach by Helfand and Wertheimer~\cite{HW}. Unfortunately, a
consistent solution of these problems at any temperature between 0 and
$T_c$ requires much more detailed knowledge about the Fermi
surface. For example, to reach a satisfactory convergence
Miranovi\'c~\etal~\cite{kogan2} were forced to introduce $11$
different parameters characterizing the Fermi surface and electron
interaction. It is clear, that our real knowledge of the Fermi surface
is too poor for such a sophistication. Dahm and Schopohl~\cite{dahm}
applied a simplified model of the Fermi surface as consisting of a
torus and cylinder characterized by 4 parameters only and assumed a
plausible variational procedure introducing one more parameter. As it
could be expected from the results by Miranovi\'c~\etal, the number of
parameters is too small to ensure a reasonable precision. Indeed, a
satisfactory agreement with the experiment in Ref.~\cite{dahm} is
reached at the expense of a rather exotic choice of parameter. Summing
up, the magnetic properties can not be described by such an elementary
theory as the described above two-band BCS model and require much more
sophisticated approach even in the weak coupling approximation.

Let us discuss why this simplified theory  works so well. Let us
start from the assumption supported by experiments that the gap
does not vary within each band. The in-band isotropy of the gap
could be a result of sufficiently strong in-band scattering. At
the scattering time $\tau\sim 10^{-14}$ s, i.e. at the residual
resistance larger than $10^{-5}\,\Omega\mathrm{cm}$, the energy
gap becomes isotropic. However, the ratio of the gaps for
different bands still remains bigger than 2 indicating that the
inter-band scattering must be much weaker. It should be emphasized
that it is the density of states which becomes isotropic, whereas
the order parameter remains anisotropic unless the Ioffe-Regel
limit $\tau\varepsilon_F\sim 1$ of scattering rate is
reached~\cite{PP}. The tunnelling experiment measures just the
density of state.

Second question is why the weak-coupling model gives so high
accuracy. Two different aspects must be enlightened. First, the
separability of variables for the order parameter, even in the
framework of the weak-coupling approximation, has the precision of
of the weak coupling constant, i.e. $(\ln\frac{\Delta}
{\omega_D})^{-1}\sim 0.3$. For the case of the two-band model such
a crude estimate can be checked more accurately by a direct
solution of the nonlinear matrix equation for the energy gap. It
has a following form:
\begin{equation}
\Delta_i=\sum_j
V_{ij}c_j\Delta_j\left[\frac{1}{\lambda_+}-f(\beta\Delta_j)\right]{\Delta_j}\,,
\label{non-linear}
\end{equation}
where $i,j$ take values $\sigma$, $\pi$ and
$f(x)=\int_{=\infty}^{\infty}\left(\frac{\tanh u}{u}-
\frac{\tanh\sqrt{u^2+x^2}}{\sqrt{u^2+x^2}}\right)du$. Its solution
can be found as a superposition of two normalized eigenstates of
the corresponding linear equation:
$\Delta_j=Q_{+}\Psi_{+j}+Q_{-}\Psi_{-j}$. In our calculations we
used only one of them, $\Psi_{+}$ corresponding to the larger
eigenvalue $\lambda_{+}$. Such an approximation is justified when
the second eigenvalue $\lambda_{-}$ is much less than
$\lambda_{+}$, even if $\lambda_{+}$ is not very small. Indeed,
the symmetrized matrix $\tilde{V}$ with matrix elements
$\tilde{V}_{ij}=\sqrt{c_ic_j}V_{ij}$ can be represented as
$\tilde{V}=\lambda_{+}|+\rangle\langle
+|+\lambda_{-}|-\rangle\langle -|$. This representation shows
that, at $\lambda_{-}=0$ the operator $\tilde{V}$ is separable,
and the solution of non-linear equation (\ref{non-linear}) is
factorizable: $\Delta=Q(T)\Psi_{+}$. The equation $\lambda_{-}=0$
is equivalent to ${\rm Det}
V=V_{\sigma\sigma}V_{\pi\pi}-V_{\pi\sigma}^2=0$. Though such a
fine tuning of parameters seems improbable, our numerical
calculations demonstrate that the ratio $\lambda_{-}/\lambda_{+}$
and the thermal variation of of the ratio
$\Delta_{\pi}/\Delta_{\sigma}$ remain small (about 3\%) even at
${\rm Det} V/(c_{\sigma}V_{\sigma\sigma}+c_{\pi}V_{\pi\pi})\sim\pm
0.2$. Thus, the experimental facts seem to indicate that one of
the two eigenvalues is significantly smaller than another. Such a
situation occurred earlier in a band calculation for
high-$T_c$-superconductors \cite{mishonov-band}.

The second aspect mentioned in the preamble is that the BCS
approximation itself has a low precision and should be substituted
by the Eliashberg formalism. The numerical calculations by
Golubov~\etal~\cite{golubov} indicate that the Eliashberg weight
functions are very small in a broad range of low energy and has
rather sharp peaks in the range of 800--1000~K. This is an unusual
situation. Leavens and Carbotte~\cite{carbotte} considered an
extended Eliashberg weight function $\alpha^2 F(\omega )$ centered
at values $\omega\sim \omega_0$ much larger than the
superconducting energy gap $\Delta (0)$. They argued on the basis
of numerical calculations that in this case the function $\Delta
(\omega )$ varies very weakly at $\omega <\omega_0$ and then
rapidly changes sign. They even modeled $\Delta (\omega )$ by the
step function. Their arguments seem to be correct for the
considered case as well. Then it is obvious that by integrating in
the range of high frequency, it is possible to obtain the BCS-like
equations with a renormalized, not small interaction between
electrons with momenta on the Fermi surface. Though such an
explanation is plausible, further study of the Eliashberg equation
with a model weight is highly desirable.

\acknowledgments
We are thankful to Dr. A. Junod and to Dr. R. Gonnelli for sending us
original experimental data of their works. This work was supported by
NSF under the grants DMR-0321572 and DMR 0103455.

\end{document}